Excerpt from *Introduction to Cyber-Warfare: A Multidisciplinary Approach*
*(Shakarian, Shakarian, and Ruef)*

# Cyber Attacks and Public Embarrassment:

# A Survey of Some Notable Hacks

**Jana Shakarian, Paulo Shakarian, and Andrew Ruef**

We hear it all too often in the media: an organization is attacked, its data, often containing personally identifying information, is made public, and a hacking group emerges to claim credit. In this excerpt, we discuss how such groups operate and describe the details of a few major cyber-attacks of this sort in the wider context of how they occurred. We feel that understanding how such groups have operated in the past will give organizations ideas of how to defend against them in the future.

*Hacking Group Modus Operandi: the Case of Anonymous*

> "I think what is, important about Anonymous is that it works, I don't know how it works but it works"
> Australian Anon quoted in NineMSN, 05/08/2008[1]

A video, mostly published on YouTube, announcing a new campaign kicked off many projects in the past.[2] Additionally, every "operation" ("AnonOp") is declared beforehand on Twitter (@YourAnonNews, @Anon_Central), tumblr (youranonnews.tumblr.com) and first-hand Anonymous websites, such as anonnews.org and whyweprotest.net. Other commonly used outlets are Pastebin, Pirate Bay[3] and Facebook (Anonymous News Network).[4] Internet Relay Chats (IRC) and image boards, Internet forums, as well as wikis serve the purpose of online communication and coordination.[i] These as well as the partyvan.info[5] hosted raid boards and IRCs as well as those run by AnonOps (anonops.ru,

---

[i] Especially 4chan, 711 chan, Encyclopedia Dramatica, IRC channels, YouTube, and (supported) blogs (e.g. anonops.blogspot.com (by invitation only), Legion News Network and AnonNews.org as news outlets, whyweprotest.net (Anonymous supported website originally in support of operation Chanology, but expanded to service the organization of protest initiatives and campaigns, grass root activism), and Anonywebz.com (Anonymous tool network), social networking services like Twitter and Facebook are used to organize real world protest

anonops.net) are apparently used to garner support for ideas from individual Anons. The websites of collaborating hacktivist groups, too, may explain the mission and extend the opportunity to participate to the visitor.[6] One *chan* user might post a call to arms where s/he perceives something is wrong or has potential of being fun. If enough other users agree a date is set on which the operation is launched.[7] If the initial recruiting process was unsuccessful, nothing will happen. Factors that might motivate an Anonymous-attack (i.e. further the recruitment process to an individual idea) are the *lulz-potential*[i], perceived arrogance of the potential target, and perceived Internet censorship or other ethical or moral challenges.[8] The latter may be secondary or even afterthought for the mere justification of the hacks for the public eye and the decisive moment is the actual possibility to achieve the desired results (although the ensuing real world consequences may not necessarily have been anticipated fully).[9]

The most often employed tactic in Anonymous & Co.'s campaigns is the distributed denial of service attack (DDoS). As with the some of the other DDoS attacks discussed in previous chapters, the DDoS attacks of Anonymous rely on DDoS tool provided to the hacktivist participants. These tools were somewhat more advanced than the software used by Russian hacktivists in the Georgia cyber campaign (chapter 3), but unlike the "voluntary botnet" used by pro-Israel supporters against Hamas (chapter 4) they do not allow a central C&C total control over the user's computer. The creation of such tools allows Anons to participate with virtually no hacking skills – while still contributing their computing resources. The most widely known software of this sort in use by Anonymous includes the Low Orbit Ion Cannon (LOIC) and the RefRef web script[10]. The latter of which was meant to replace the former in September 2011 since LOIC retains the IP-address of the senders, which lead to numerous arrests of Anons.[11] The RefRef web script is a command-line based Java site which exploits SQL and Javascript vulnerabilities in order to exhaust the server supporting the website.[i] The Low Orbit Ion Cannon is a legitimate tool to stress-test a web-application by sending a large amount of requests to the webpage to see, if it can handle it ("bandwidth raep"). It was written in C#, is available in three versions (manual, server-controlled, and in JavaScript) and can be found on any major open source code repository. Individual users download the application and opt their computer to become part of the botnet which is employed in Anonymous' DDoS-attacks; one form of membership is thus the mere contribution of computing power to Anonymous' resources. LOIC also features a connection to an IRC channel in order to coordinate the load of packets send from the individual computers. Unbeknownst to a large number of participants in illegal DDoS-attacks may be the fact that LOIC does not encrypt or hide the IP-addresses of the senders. The target computer's website logs all IP-addresses of incoming requests which in turn can be traced back to a single computer. [12]

---

[i] According to Adrian Crenshaw ("Crude Inconsistent Threat") this is what appears to draw the most supporters

Another prevalent Anonymous tactics requiring somewhat more hacking skills, but less computing resources is the SQL injection, which exploits weaknesses in the database of a website – previously described in chapter 3. In its attack on HB Gary Federal and its head, Aaron Barr, a handful of Anonymous and affiliated hackers gathered compromising documents and emails this way. Captured data usually is "dumped", i.e. published on Pastebin or most recently on Anonpaste. More rarely, the obtained data is used for In Real Life (IRL) pranks – the staging of real world events such as unwanted pizza delivery or using social engineering to have a SWAT team called to the respective residence ("swatting").[13] Other ends through the mean of SQL injection include the defacement of websites, in which the original content is substituted by pictures or messages of the intruders or the hijacking and/ or redirection of the target website. In the latter cases the website would either cease to be under the control of its original authors or redirect to another site as selected by the hacker(s).

---

[i] The tool exploits the fact that most web sites save the .js file of an incoming request on their own server, since the request is still working the packets bounce back and forth on the target server, exhausting its resources (THN, 07/07/2011)

### *Targeting Governments, Corporations, and Individuals: Notable Hacks on Anonymous*

Anonymous meanwhile is featured in the media almost every week with either claiming or being found responsible for cyber-attacks against selected targets. It would not make any sense to recap a chronology of Anonymous' attacks as the resulting long list would probably have a lot of holes: the media coverage of Anonymous- acts depends either on largesse or popularity of the target.

*Habbo Hotel Raids*

Long before Anons engaged in hacks for which the collective know is famous and feared for, 4chan's Ur-Anonymous raided their targets. Protesting against Habbo's ban on African-American avatars and accusing Habbo's social moderators of racism the virtual game made a great target for raids. Habbo Hotel raids began in mid-2005 in which avatars looking like a wealthy African-American in business suit (to some reminiscent of the Samuel Jackson's Pulp Fiction character Jules) obstruct entry to the pool. This hack already signifies two important motifs for the collective action of the Anonymous we know today: perceived discrimination as well as the humungous lulz potential.[14]

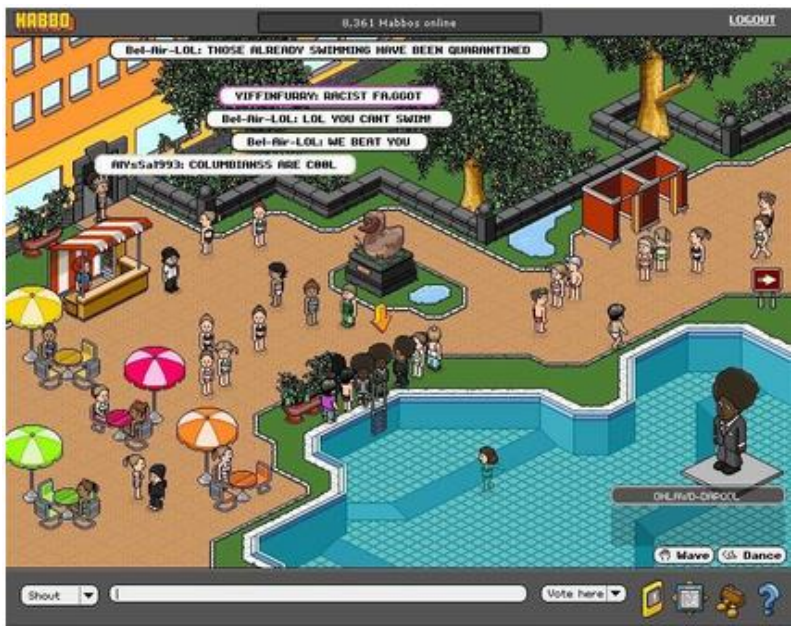

Excerpt from *Introduction to Cyber-Warfare: A Multidisciplinary Approach (Shakarian, Shakarian, and Ruef)*

> *Figure: During the Habbo hotel raid, members of 4chan protested against the ban on African-American avatars by creating avatars of African-American businessmen who blocked access to the pool. Source: KnowYourMeme[15]*

"Pool's closed" became a catch-phrase for the Habbo raids that resulted in 4chan exercising otherwise rare censorship and moderation. A year later Habbo -raids culminated in an unprecedented organized raid across four countries (UK, USA, Germany, and Australia) that resulted in the site being knocked offline. The Habbo-raids are also a precedent for the counterproductive actions that bring about what (at least part of) the momentary Anonymous force set out to prevent in the first place and which will form the core of the criticism of Anonymous years later: In the aftermath of the summer 2006 attacks Habbo Hotel administrators programmed the automatic ban of African-American avatars with afro and suit.[16]

*Internet Vigilantism*

In December of 2007 in an act of Internet vigilantism[17] Anonymous hackers aided the arrest of Canadian Chris Forcand by impersonating teenage girls[i].[18] He would subsequently be brought to court on charges related to pedophilia.[19] In summer 2012, Anonymous followed up on this initiative against pedophiles in launching the Twitter hashtag "#TwitterPedoRing" to invite pedophiles on the communication network to join. Week later 190 offenders were arrested by the U.S. Immigrations and Customs Enforcement, though no reference to Anonymous was made. In August 2012 the Anonymous Twitter account, @YourAnonNews, asked its 615,000 subscribers to report the account "@many501611" spam who had published pornographic pictures of young boys.[20] Other, instances in which the propagation of evidence of perceived injustice through the Internet resulted in real-life consequences for the perpetrator include a South Korean girl who refused to clean up after her dog on a subway and whose subsequent harassment brought her on the brink of suicide, the tactless Zhang Ya, whose inhuman remarks on the suffering of the victims of the Sichuan earthquake in 2008 eventually got her arrested, American pioneer spammer, Alan Ralsky, receiving mailbox spam in truck loads, the New York City cop, Patrick Pogan, who attacked an unsuspecting cyclist and would have gotten away with it, if it was not for the video of the incident being spread all over the Internet.[21] Internet vigilantism is neither an invention of 4chan nor the Anonymous collective nor constrained to North America.[i]

---

[i] The online threads from the perspective of the anonymous collective of b/tards are saved on 4chan's archive here: http://4chanarchive.org/brchive/dspl_thread.php5?thread_id=42828652&x=brb%20church%20-%20chris%20forcand

*Project Chanology*

Anonymous first real-world protest brought thousands donning Guy Fawkes masks onto the streets in numerous cities and countries to protest against the Church of Scientology. Churches experienced real-world pranks, like harassing phone calls, black faxes (in order to waste ink)[22], pizzas and taxis they did not order[23] as well as letters with what turned out to be harmless white powder[24]. Project Chanology was sparked by the leak of an insider video of an interview with actor Tom Cruise talking about his life in the sect.[25] The subsequent forceful attempts by the Church to retract the video were perceived as a metaphorical example for the general restrictiveness the organization is often criticized for. The initial momentum that kicked of immense headaches for the Church of Scientology was an anonymous post on 4chan/b: "it is time to do something great".[26,27] In January 2008 self-confessed Anons originating from image boards like 4chan, Partyvan.org, and 711chan amongst others officially launched "Project Chanology" by posting a video on YouTube entitled "Message to Scientology".[28] Besides the real-world protests, Anons rendered the official website of the Church of Scientology inaccessible by way of DDoS-attacks and captured and published literature for which adherents normally have to pay.[29, 30] In the course of three weeks, Project Chanology encouraged over 8,000 harassing phone calls, 3.6 malicious emails, 141 million hits against the church web sites, 10 acts of vandalism, 22 bomb threats, and 8 death threats.[31] The Internet activists employed online tools like JMeter and Gigaloader in DDoS-attacks and the exploit of computer security vulnerabilities like Cross-site Scripting.[32] Botnets and the Low Orbit Ion Cannon (LOIC) had their debut as Anonymous' tools in Project Chanology.[33, 34] Scientology responded with attempts to obtain restraining orders for their premises and identify the Anons who organized local protests. One of the exposed, Gareth A. Cales of Los Angeles reported legal threats and general harassment by Scientology members.[35] His account also includes details on the organization of the protests associated with Project Chanology. Mark Bunker and other critics of the Church of Scientology released a video encouraging the protests to remain legal and were surprised that Anons appeared to heed the advice.[36] Four years later, in February 2012, still some protesters wearing Guy Fawkes masks beleaguer Scientology churches.[37] The two-pronged approach showed effect: The resounding response to the Project Chanology not only made the organizers aware of the possibilities of the online organization of a social movement, but it also appears to have refined Anonymous' character. While the majority of the Anonymous collective prior was interested in the lulz-potential of an undertaking[38], with the protest against an enigmatic organization known for its attempts to censor public information about itself, many Anons appear to be spurred by the perceived righteousness of this quest.[39]

---

[i] Clay Shirky authored an insightful book ("Here Comes Everybody – The Power of Organizing Without Organization") on online group action, which argues that it does not only take new forms of technology, but also new forms of behavior.

*Arab Spring*

Anonymous targeted government websites of countries in North Africa and the Arabian Peninsula as well as Asia Minor that found themselves engulfed in civil uprisings in 2011 – the Western world would call it hopefully "Arab Spring". Anons used their computers and knowledge to support the protesters on the ground, whose governments in many cases had resolved to deny them access to social media or the Internet altogether. For each country, Anonymous launched a new "operation". Starting with "Operation Tunisia" and "Operation Egypt" affiliates of the collective and sympathizers took down the governments' computers with DDoS-attacks.[40]

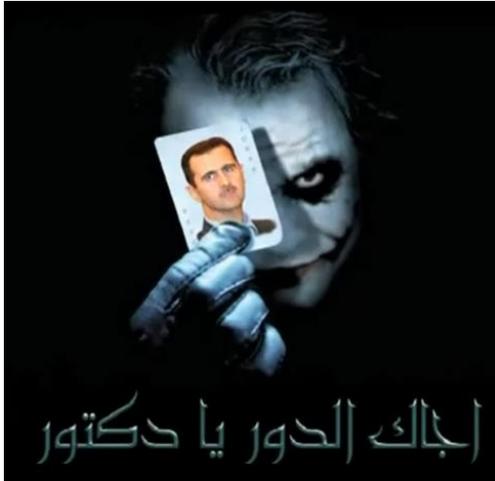

**Figure: 2011, Anonymous consequently took up the fight with the Syrian government after it clamped down violently on public protesters. Source: PLF, Operation Syria[41]**

The PLF's Commander X boasted in a chat-interview that his organization kicked off Tunisia's "Jasmine Revolution" by stealing compromising information on the government and leaking it to WikiLeaks.[42,43] Other media outlets stress the dire living conditions in Tunisia, amongst others the high unemployment and inflation as well as the lack of political freedom as causes taken to Tunisian streets after the self-immolation of Mohamed Bouazizi on December 17, 2010.[44,45] The press acknowledged Anonymous' support of the protesters on the ground by taking down at least eight government websites with DDoS-attacks (including websites of the president, prime minister, the ministries of foreign affairs and industry, as well as the stock exchange)[46] and defacing some of them[47]. Anonymous also provides means and knowledge to activists on the ground to conceal their online identity in order to prevent prosecution and coordinate their activities with the events on the ground.[48] As Commander X stated in regard to their participation in various episodes of the "Arab Spring", besides the admittedly mostly negative attention Anonymous and its collaborators earn with each campaign, the goals are to deny the targets (here: governments) lines of communication and to "encourage the protesters on the ground".[49]

The initial motivation to join in the uprising for one Anon was the Internet censorship as exercised by the Tunisian government.[50] Subsequently a host of Anons identified with the struggles of the revolutionaries in the respective countries and sought to alleviate their plight by doing whatever they could to provide them with avenues for information exchange[51] while at the same time denying them to their opponents. The Tunisian governments attempt to identify political activists by phishing online for their social networking or email accounts[52] was countered by a subdivision of Anonymous, the *Internetfeds*. A greasemonkey script made available to the Tunisian protestors would override the governments' spyware.[53] They also translated and distributed information on how to conceal IP-addresses and other techniques to obscure individual identities in French and Arabic.[54] Anonymous' involvement did not stop with the ouster of Zine el Abidine Ben Ali. The new Tunisian Anons proceeded to use online tools to fight alongside the new moderate Tunisian government against the pressure it experienced from Islamist fringe groups.[55] To circumvent Internet censorship in Egypt, Anonymous aided in the restoration of proxies and mirrors in order to guarantee information flow to benefit the revolutionaries.[56] The Egyptian authorities had blocked twitter.com and other social media apparently in order to aggravate communication between protestors.[57] Anons targeted websites of the Egyptian government and ministries with DDoS-attacks.[58] In both operations the trickster-natured among the Anons still found pleasure in real-world pranks, like flooding the respective embassies with unwarranted pizza-orders.[59] In August 2011, Anonymous consequently took up the fight with the Syrian government after it clamped down violently on public protesters. The website of the Syrian Ministry of Defense was defaced with an encouraging message to the Syrian people.[60]

Besides the above mentioned, the governments of Yemen[61,62], Algeria[63], Zimbabwe[64], and Italy[65] also drew the ire of the online collective, which targeted associated websites with more or less extensive DDoS-attacks.[66] The predecessor to these diverse operations associated with the "Arab Spring" could

Excerpt from *Introduction to Cyber-Warfare: A Multidisciplinary Approach*
*(Shakarian, Shakarian, and Ruef)*

perhaps be found in Anonymous' engagement in Iran's post-election crises in 2009. Back then Anons intercepted the propagation of hit lists depicting protestors distributed by pro-government groups.[67]

*HBGary Federal and Aaron Barr*

The InternetFeds, an AnonOps spin-off of skilled black hat hackers engaged in what appears to be a "personal" vendetta on Aaron Barr, HBGary's CEO, who had claimed to have identified key actors after infiltrating Anonymous.[68] After mocking Barr online and obtaining all his passwords via SQL injection the hackers were amused to have him "watch" how they hijacked his email and all Internet social media and stole data from internal company emails he had sent to his coworkers.[69] DDoS attacks brought down hggary.com and hbgaryfederal.com.[70] The latter was breached using SQL injection, which exploited a security flaw in the custom-made content management system (CMS) the company had employed.[71] So called rainbow tables appeared to have immensely alleviated the quest to garner the hashed passwords of Aaron Barr and COO Ted Vera.[72] Barr had apparently re-used one password over and over, "kibafo33", which allowed access to other company email-inboxes since Barr enjoyed administrator rights. [73] Ultimately this led to compromising HBGary's founder, Greg Hoglund's rootkit.com website by way of hijacking his email account. The hackers then impersonated Hoglund in an email-conversation with an associate in order to obtain the last bits of information necessary to gain control over rootkit.com.[74] Besides defacing the website, the intruders also resolved to publish the user database.[75] Ted Vera's repeatedly used password enabled the hackers to access the support server that hosted the shell accounts of many HBGary employees and which enabled them to upgrade Vera's personal account through the exploitation of a personal escalation vulnerability, which gave the hackers full access to HBGary's system.[76] Gigabytes of data were now immediately at the hand of the intruders. More than sixty thousand company emails were published on the Pirate Bay file-sharing site.[77] Another case apparently motivated by revenge was the hack into Sony's online Playstation store and the DDoS-attack against the Playstation website.[78] Sony had sued PS3-hacker George Hotz (aka "GeoHot"), who had become famous for unlocking various iPhone versions[79] before he managed to gain administrator-rights to the entire system memory and the processor of the PS3 system in early 2011.[80] An abrupt settlement in the case was reached in April 2011 with neither side disclosing the full terms, but on the part of the phreaker/hacker was the deletion of all his information on the PS3-hack from the Internet.[81] Hotz was subsequently hired by Facebook, but found himself just months later with Lady Gaga's start-up social networking site, Backplane.[82]

*Straightforward Operations*

Anonymous' goals become apparent when looking at those campaigns that aim at institutions that are either held responsible or perceived to embody the objectionable with DDoS-attacks. In some cases sensitive material is published after the successful intrusion of hackers into the target's system. A prominent example is the charge under which attacks against copyright companies were launched. The effort was later extended to also aim at companies that ceased business relation with the whistle-blower, WikiLeaks, after the publishing of sensitive U.S. diplomatic cables (Operation Payback).[83] In this instance, Anonymous officially became a supporter of WikiLeaks, its co-founder, Julian Assange as well as Bradley Manning, a U.S. soldier who had leaked confidential information on America's military campaign in Iraq.

The term "anti-security" refers to a movement conceived in the late 1990s to counteract the cyber-security industry's tactic of provocatively exploiting vulnerabilities in order to increase sales[84]. Apparently, Anonymous joined by LulzSec[85] takes the viewpoint that governments employ similar tactics to justify legislation that monitors Internet behavior and allows censorship and surveillance (Operation AntiSec). In its quest to safeguard online privacy, the hacktivists encouraged the defacement of government websites and leaking of data obtained through breaches into the systems government organization, banks and other institutions deemed to profit from infringements on user privacy.[86] Other digital rights-related reasons for hacks under the AntiSec banner are racial profiling, copyright laws and the War on Drugs[87]. The attack against the British SOCA presented the debut of AntiSec, followed by the publication of material from the Cyberterrorism Defence Initiative Sentinel program[88], and the release of private information apparently obtained through SQL-injection from sources related to the governments of Brazil, Zimbabwe, and Tunisia. So far Anonymous attacked U.S. defense contractor Booz Allen Hamilton, U.S. Central Command, U.S. Special Operations Command, U.S. Marine Corps, and the U.S. Air Force in a spree dubbed "Military Meltdown Monday".[89] The December 2011 Stratfor-hack where subscribers' credit card information and emails were stolen and later published on WikiLeaks was also perpetrated by the LulzSec member(s) now working under the Anonymous banner[90].

Other campaigns launched under the Anonymous banner include Operation Vendetta[91] launched after a wave of arrests of alleged Anonymous-affiliated hackers in March 2012 to help Anons escape the authorities. This campaign is the second one to date that entails a major real-world component in that it aims to set up safe-houses and an "underground railroad"[92] to lead sought-after hackers to countries which don't have extradition treaties with the prosecuting authorities. A popular hacktivist who apparently benefitted from this operation is Christopher Doyon aka Commander X, whose People's Liberation Front had joined forces with Anonymous in this campaign again. After having been arrested in Santa Cruz, California, as the "homeless hacker"[93] who protested against a recent county ban on outdoor sleeping by launching DDoS-attacks[94], Doyon reportedly found an interim sanctuary in Canada[95].

**Software for the Legion: Anonymous Products**

In early 2012 Anons engaged in the creation of software. Partially in protest to existing applications, such as the social music platform, *AnonTune*, and partially in the quest of opening new avenues beyond hacking as in the launch of its operating system, *Anonymous OS live*. Besides the file-sharing site, AnonPaste, the collective also launched a WikiLeaks-like site in an attempt to control and deepen the public impact of their hacking activity[96]. Both platforms were meant to serve as data dumps.

*AnonTune*

In April 2012 Anonymous launched a social music platform that provides streaming songs from third party users (e.g. YouTube) for users to compile into playlists and share. Initiated by online discussions the ambitions of an individual, who created and uploaded a prototype, generated the technical support necessary to render AnonTune capable of handling large requests aside from the user interface[97]. It is planned that AnonTune will grant its users anonymity, shielding them from prosecution and copyright law suits while offering a service better organized and less expensive than similar platforms.[98] Upon its launch AnonTune will be a project borne purely of the Internet. Once more, the creators of this online forum must not have ever been in personal physical contact and may even be spread out over different continents.

*AnonPaste*

After the operators of PasteBin, hitherto Anonymous' favorite forum to publish captured data, was found to be compromised, Anonymous affiliates associated with Operation Anti-Sec together with the People's Liberation Front launched its own data-sharing website, AnonPaste[i], in April 2012.[99] Jeroen Vader, owner of PasteBin, announced an increase in censorship of the site[100]. He further admitted that at times he shares his logs -a register of the IP-addresses of users - with law enforcement agencies, which drew immediate recognition on the part of Anonymous, which introduced its new website with the following comment:

> "As a recent leak of private emails show clearly, Pastebin is not only willing to
> give up IP addresses to governments – but apparently has already given many
> IPs to at least one private security firm. And these leaked emails also revealed a
> distinct animosity towards Anonymous"
> (RT.com 23 Apr 2012)

---

[i] Anonpaste.org, last accessed 05/12/2012 – no entries

AnonPaste allows the user to determine when the post is to be deleted at a time increment of his or her choosing. The creators of this new haven for captured data promote a secure forum without advertisements, censorship or connection logs, and which retains only encrypted data.[101] AnonPaste was registered in the New Zealand –administered Tokelau (.tk) which allows everybody to register a free domain. It relies solely on donations as advertisements are not featured.[102]

When WikiLeaks sought to verify the data captured during the intrusion into the intelligence company, Stratfor in December 2011, part of Anonymous was frustrated with the slow release of their bounty and launched Par:AnoIA (Potentially Alarming Research: Anonymous Intelligence Agency). It also appears to be an attempt at solving a problem high-volume websites are struggling with everywhere: the sensible organization of humungous amounts of data.[103] The Anons who run the page say their sole task is to present the data others from the Anonymous collective have hacked in a usable format.[104] At the time of writing neither AnonPaste nor Par:AnoIA were workable websites.

*Anonymous-OS 0.1/ Live*

The Ubuntu 11.10 -based operating system released for "education purposes"[105], Anonymous OS Live allegedly performed security checks on websites testing for password-security and simulating DDoS-attacks.[106] As default search engine it used DuckDuckGo and it came with a host of pre-installed applications, which included Tor, Slowloris and AnonymousHOIC.[107] In its initial news release The Hacker News warned that a genuine source of the operating system is not apparent and that it could be "backdoored" by law enforcement or affiliated hackers.[108] Accompanying the software a Tumblr and About page provided news and updates and the disclaimer that the user - not the developers - is responsible for any illegal use of the operating system.[109, 110]

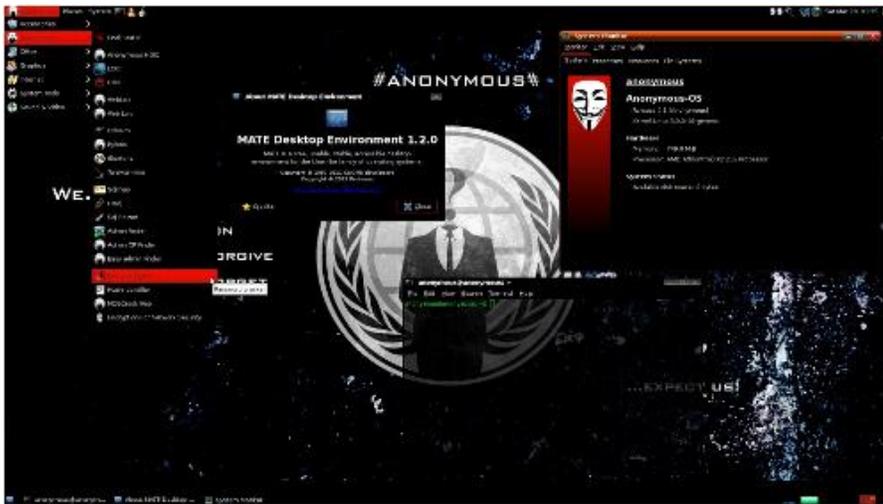

*Figure: Screen-shot of the Anonymous Operating System - essentially a Linux distribution. Source: Softpedia, 03/16/2012*[111]

In the four days following its release the operating software was downloaded over twenty-six thousand times and enjoyed user ratings of 62%.[112] The free application was quickly marred by doubts about its true origin and the website hosting its free download, SourceForge, eventually closed the project site and removed the downloads from its server.[113] Official Twitter accounts, AnonOps, YourAnonNews, and AnonNewsSec, warned it was "wrapped in Trojans"[114] while its creators denied it was spyware.[115, 116]

In July 2012 an effort to represent itself in the media world without the bias other news outlets[117] are accused of carrying took forms in "AnonPR" (anonpr.net, complete with newsletter subscription) and "Anonymous Analytics" (anonanalytics.com, featuring "research" and a mirror of media reports on Anonymous).

*Chapter Summary*

This chapter represents the attempt to outline what the hitherto most notorious hacker collective, Anonymous, might look like on a whiteboard. Publically available hints towards its structure had been followed and explored: the possible initial "directorate" recruited from an idealist veteran hacking group as well as the better known online seedbed, a mingling place for contempt and counterculture. Whatever the beginnings, Anonymous is driven by a set of motivations largely shared by those whose actions are most visible: freedom of information and the right to online privacy. There are a certain number of online arenas which are used to decide upon, plan and organize a hack – mostly using SQL-injection or a related, basic hacker skill that grants access to target systems. Personal, confidential and otherwise compromising data is later dumped on one or more popular file-sharing sites mainly to give evidence of the intrusion. Mostly after capturing sensitive data, but not requiring this more or less clandestine step a DDoS-attack is used to render the target website inaccessible. In some cases website defacement prominently displayed the reason and goals of Anonymous' interference. Over its history the collective experienced dedicated functional spin-offs (e.g. LulzSec) or ideological derivatives who sought to be more "pure" (e.g. MalSec) whose history and demise is antidote as well as defining element of Anonymous' structure. The political hacktivists also encountered like-minded collaborators (e.g. PLF), who probably helped emphasize this facet of the collective. Finally, this chapter sought to display representative motivations, modi operandi, and tools through the brief description of select hacks. In the wake of Anonymous' refocusing or just the skilled public relation work of a few Anons, the chapter concludes with examples for the nascent stage of products: a music portal, two dumping sites and a highly controversial, quickly retracted operating system.

Anonymous might be more the history of hacking exploits than a social structure, but in the role of Sabu alone it contradicts its claims of being a legion of countless "everybodies." Sabu organized, motivated and lead many of the highly popularized hacks in 2011. His absence after the March 2012 arrests was noticeable and it forced LulzSec to reinvent itself. The global wave of arrests in 2011 and especially those in early 2012[118] are perceived by some as crippling for the Anonymous collective[119], which can only be the case for an organization that is hierarchically structured. Detentions can have effects on the activity-level of a group only, if the arrested individuals were crucial in the organization of the activities. Whether there is or was a handful of people conceiving and directing the politically aware and active Anonymous, the collective appears to be much more than a loosely-knit organization now. The collective action presented in its every hack represents an enormous challenge not only to the reader who seeks to wrap her head around this virtual phenomenon, but also to the social scientist. Initially self-proclaimed members of the collective without any real-world connection to fellow hacktivists, political Anons may form real local groups with regular meetings and faces, real names. In the virtual meeting spaces (user) name recognition still applies and allows for virtual groups and Anonymous-spin offs to be formed. But the numerical majority of the collective remains elusive with many different levels of possible engagement ranging from sympathizers to participants in DDoS-attacks. The world-spanning virtual social network, which conceives, decides, plans, and organizes hacking exploits is what and who Anonymous really is: a number of IRC-channels, blogs and message boards accessed from (at least) several hundred thousand devices all around the globe. 4chan might have been its cradle, but so far it seems Anonymous has risen beyond this tactless, seedy playground with its bored opportunists, tricksters and hustlers.

So far it seems the political activists have by and large conquered the movement, though it cannot completely abandon its trickster nature. The Janus-faced character of the collective is reflected in its every aspect from activity, to the motivation and the understanding of itself is due to the myriad of individuals who have used the Anonymous platform for very different reasons. In a handful of interviews some self-proclaimed Anons try to fixate an image for the collective, but due to the elusiveness of its membership it will be difficult to instill and maintain. The political hactivists the Anonymous collective depend on its favorable depiction in the media since the nature of its preferred modus operandi, the employment of the low-orbit ion cannon in DDoS-attacks, hinges on a large number of volunteers. The advance in new alleyways with the launches of the Anonymous operating system, the music sharing website as well as the data dump sites, albeit with different levels of success, help diversify the collective, but are also evidence of the lack of guidelines for members, that could serve as identity markers.

*Suggested Further Reading*

Parmy Olson's 2012 book *We Are* Anonymous – *Inside the Hacker World of LulzSec, Anonymous, and the Global Cyber Insurgency*, is a very readable synergy of documentary and novel that explores aspects of the *Anonymous* collective by following the exploits of a fictitious hacker. Some of Olson's story is mirrored in insider Cole Stryker's account of the legendary *4chan* message board in *Epic Win for* Anonymous – *How 4chan's Army Conquered the Web* (2011).

For a more, general perspective on how the Internet has changed social movements, we recommend Clay Shirky's 2008 *Here Comes Everybody- The Power of Organizing Without Organizations*. Finally, one of the best known early computer espionage stories is Clifford Stoll's classic *The Cuckoo's Egg – Tracking a Spy Through the Maze of Computer Espionage*, which provides a first-hand account of the KGB-hack. This was later made into a television documentary entitled *The KGB, the Computer, and Me*.

Excerpt from *Introduction to Cyber-Warfare: A Multidisciplinary Approach*
*(Shakarian, Shakarian, and Ruef)*

[1] NineMSN, 05/08/2008, "The Internet Pranksters who Started a War", URL: http://news.ninemsn.com.au/article.aspx?id=459214

[2] Baltimore City Paper, 21 Jan 2008, "Serious Business", URL: http://www2.citypaper.com/columns/story.asp?id=15543)

[3] IT World, 13 Feb 2011, "That new Facebook Friend Might Just Be a Spy", URL: http://www.itworld.com/internet/136830/that-new-facebook-friend-might-just-be-a-spy

[4] Times of India, 11 Feb 2012, "Anonymous hacks CIA website", URL: http://articles.timesofindia.indiatimes.com/2012-02-11/security/31049746_1_cia-site-cia-website-anonymous

[5] Irongeek (Adrian Crenshaw), undated, "Crude, Inconsistent Threat: Understanding Anonymous", URL: http://www.irongeek.com/i.php?page=security/understanding-anonymous

[6] Peoples Liberation Front, Public Campaigns, URL: http://peoplesliberationfront.net/campaigns.html

[7] RFE/ RL, 3 Mar 2012, "What Is 'Anonymous' And How Does It Operate", URL: http://www.rferl.org/content/explainer_what_is_anonymous_and_how_does_it_operate/24500381.html

[8] Irongeek (Adrian Crenshaw), undated, "Crude, Inconsistent Threat: Understanding Anonymous", URL: http://www.irongeek.com/i.php?page=security/understanding-anonymous

[9] Cognitive Dissidents (Josh Corman, Brian Martin), 29 Dec. 2011, "'Building a Better Anonymous' Series, Part II: Fact vs Fiction", URL: http://blog.cognitivedissidents.com/2011/12/29/building-a-better-anonymous-series-part-2/

[10] ZDNet, 17 Feb 2012, "Anonymous Launches 'Operation Global Blackout', aims to DDoS the Root Internet server", URL: http://www.zdnet.com/blog/security/anonymous-launches-operation-global-blackout-aims-to-ddos-the-root-internet-servers/10387

[11] The Hacker News, 7 July 2011, "#RefRef – Denial of Service (DDoS) Tool Developed by Anonymous", URL: http://thehackernews.com/2011/07/refref-denial-of-service-ddos-tool.html (last accessed 02/21/2012)

[12] Information Week, 12 Dec 2010, "WikiLeaks Supporters Download Botnet Tool 50,000 Times", URL: http://www.informationweek.com/news/security/attacks/228800161

Excerpt from *Introduction to Cyber-Warfare: A Multidisciplinary Approach*
*(Shakarian, Shakarian, and Ruef)*

[13] Irongeek (Adrian Crenshaw), undated, "Crude, Inconsistent Threat: Understanding Anonymous", URL: http://www.irongeek.com/i.php?page=security/understanding-anonymous

[14] Knowyourmemes, April 2012, "Pool's Closed", URL: http://knowyourmeme.com/memes/pools-closed

[15] Knowyourmemes, April 2012, "Pool's Closed", URL: http://knowyourmeme.com/memes/pools-closed

[16] Knowyourmemes, April 2012, "Pool's Closed", URL: http://knowyourmeme.com/memes/pools-closed

[17] Knowyourmemes, May 2012, "Internet Vigilantism", URL: http://knowyourmeme.com/memes/subcultures/internet-vigilantism

[18] Cracked, 23 March 2009, "Eight Awesome Cases of Internet Vigilantism", URL: http://www.cracked.com/article_17170_8-awesome-cases-internet-vigilantism_p2.html#ixzz1RZFPycz6

[19] IB Times, 24 Oct. 2007, "Anonymous Takes on Child Pornography Sites", URL: http://www.ibtimes.com/articles/236744/20111024/anonymous-operation-darknet-child-pornography.htm

[20] Mashable, 9 Aug 2012, "Anonymous Shuts Down Alleged Twitter Pedophile", URL: http://mashable.com/2012/08/09/anonymous-twitter-pedophile/

[21] Cracked, 23 March 2009, "Eight Awesome Cases of Internet Vigilantism", URL: http://www.cracked.com/article_17170_8-awesome-cases-internet-vigilantism_p2.html#ixzz1RZFPycz6

[22] FoxNews, 25 Jan. 2008, "Hackers Declare War On Scientology", URL: http://www.foxnews.com/story/0,2933,325586,00.html

[23] WirtschaftsWoche (German language magazine), 23 July 2012, "Wie Anonymous Scientology in die Knie Zwang", URL: http://www.wiwo.de/technologie/digitale-welt/hackernetzwerk-wie-anonymous-scientology-in-die-knie-zwang/6908658.html (excerpt of Parmy Olson's 2012 book on Anonymous)

[24] The Guardian, 3 Feb. 2008, "Hackers Declare War On Scientologists Amid Claim of Heavy-Handed Cruise Control", URL: http://www.guardian.co.uk/technology/2008/feb/04/news

[25] L.A. Weekly, 4 Feb 2009, "My Date with Anonymous: A Rare Interview with the Elusive Internet Troublemakers", URL: http://www.laweekly.com/2009-02-05/columns/my-date-with-anonymous-a-rare-interview-with-the-illusive-internet-troublemakers/

Excerpt from *Introduction to Cyber-Warfare: A Multidisciplinary Approach*
*(Shakarian, Shakarian, and Ruef)*

[26] Jacobson, Jeff, undated, "We Are Legion: Anonymous And The War On Scientology", URL: http://www.lisamcpherson.org/pc.htm

[27] WirtschaftsWoche (German language magazine), 23 July 2012, "Wie Anonymous Scientology in die Knie Zwang", URL: http://www.wiwo.de/technologie/digitale-welt/hackernetzwerk-wie-anonymous-scientology-in-die-knie-zwang/6908658.html (excerpt of Parmy Olson's 2012 book on Anonymous)

[28] KnowYourMemes, Feb. 2012, "Project Chanology", URL: http://knowyourmeme.com/memes/events/project-chanology

[29] Jacobson, Jeff, undated, "We Are Legion: Anonymous And The War On Scientology", URL: http://www.lisamcpherson.org/pc.htm

[30] National Post, 26 Jan. 2008, "Online Group Declares War on Scientology", URL: http://web.archive.org/web/20080128145858/http://www.nationalpost.com/news/canada/story.html?id=261308

[31] Jacobson, Jeff, undated, "We Are Legion: Anonymous And The War On Scientology", URL: http://www.lisamcpherson.org/pc.htm

[32] WirtschaftsWoche (German language magazine), 23 July 2012, "Wie Anonymous Scientology in die Knie Zwang", URL: http://www.wiwo.de/technologie/digitale-welt/hackernetzwerk-wie-anonymous-scientology-in-die-knie-zwang/6908658.html (excerpt of Parmy Olson's 2012 book on Anonymous)

[33] WirtschaftsWoche (German language magazine), 23 July 2012, "Wie Anonymous Scientology in die Knie Zwang", URL: http://www.wiwo.de/technologie/digitale-welt/hackernetzwerk-wie-anonymous-scientology-in-die-knie-zwang/6908658.html (excerpt of Parmy Olson's 2012 book on Anonymous)

[34] CNet, 25 Jan. 2008, "Technical Aspects of the DDoS Attacks Upon the Church of Scientology", URL: http://news.cnet.com/8301-10789_3-9858552-57.html

[35] LAist, 23 March 2008, "Church of Scientology Strikes Back – Anonymous Responds", URL: http://laist.com/2008/03/23/church_of_scien.php

[36] L.A. Weekly, 17 March 2008, "'Anonymous' vs. Scientology: Group Targets 'Church' Headquarters", URL: http://www.laweekly.com/2008-03-20/news/8220-anonymous-8221-vs-scientology/2/

[37] Mancunian Matters, 13 July 2012, "Manchester Anonymous to Hold Anti-Scientology Protest Following Cruise-Holmes Split", URL: http://mancunianmatters.co.uk/content/13074409-manchester-anonymous-hold-anti-scientology-protest-following-cruise-holmes-split

Excerpt from *Introduction to Cyber-Warfare: A Multidisciplinary Approach (Shakarian, Shakarian, and Ruef)*

[38] NineMSN, 8 May 2008, "The Internet Pranksters Who Started A War", URL: http://news.ninemsn.com.au/article.aspx?id=459214

[39] WirtschaftsWoche (German language magazine), 23 July 2012, "Wie Anonymous Scientology in die Knie Zwang", URL: http://www.wiwo.de/technologie/digitale-welt/hackernetzwerk-wie-anonymous-scientology-in-die-knie-zwang/6908658.html (excerpt of Parmy Olson's 2012 book on Anonymous)

[40] Atlantic Wire, 28 Dec. 2011, "The Hacks That Mattered in the Year of the Hack", URL: http://www.theatlanticwire.com/technology/2011/12/hacks-mattered-year-hack/46731/

[41] Peoples Liberation Front, Public Campaigns, "Operation Syria", URL: http://peoplesliberationfront.net/campaigns.html

[42] IT World, 18 Feb 2011, "A Conversation with Commander X", URL: http://www.itworld.com/internet/137590/conversation-commander-x?page=0%2C0

[43] Al Arabiya, 15 Jan 2011, "Wikileaks Might Have Triggered Tunis' Revolution", URL: http://www.alarabiya.net/articles/2011/01/15/133592.html

[44] Al Jazeera, 26 Jan 2011, "How Tunisia's Revolution began", URL: http://www.aljazeera.com/indepth/features/2011/01/2011126121815985483.html

[45] BBC News, 5 Jan 2011, "Tunisia Suicide Protester Mohamed Bouazizi Dies", URL: http://www.bbc.co.uk/news/world-africa-12120228

[46] Al Jazeera, 6 Jan 2011, "Tunisia's Bitter Cyberwar", URL: http://www.aljazeera.com/indepth/features/2011/01/20111614145839362.html

[47] Gawker, 3 Jan. 2011, "Anonymous Attacks Tunisian Government over WikiLeaks Censorship", URL: http://gawker.com/5723104/anonymous-attacks-tunisian-government-over-wikileaks-censorship

[48] Al Jazeera, 6 Jan 2011, "Tunisia's Bitter Cyberwar", URL: http://www.aljazeera.com/indepth/features/2011/01/20111614145839362.html

[49] IT World, 18 Feb 2011, "A Conversation with Commander X", URL: http://www.itworld.com/internet/137590/conversation-commander-x?page=0%2C0

[50] Al Jazeera, 6 Jan 2011, "Tunisia's Bitter Cyberwar", URL: http://www.aljazeera.com/indepth/features/2011/01/20111614145839362.html

Excerpt from *Introduction to Cyber-Warfare: A Multidisciplinary Approach*
*(Shakarian, Shakarian, and Ruef)*

---

[51] Al Jazeera, 19 May 2011, "Anonymous and The Arab Uprisings", URL:
http://www.aljazeera.com/news/middleeast/2011/05/201151917634659824.html

[52] Al Jazeera, 6 Jan 2011, "Tunisia's Bitter Cyberwar", URL:
http://www.aljazeera.com/indepth/features/2011/01/20111614145839362.html

[53] Internetfeds, 6 Jan. 2011, "Remove Tunisian Government Phishing Scripts", URL:
http://userscripts.org/scripts/show/94122

[54] Al Jazeera, 19 May 2011, "Anonymous and The Arab Uprisings", URL:
http://www.aljazeera.com/news/middleeast/2011/05/201151917634659824.html

[55] Agence France Press (AFP), 17 March 2012, "'Anonymous' Group Hacks Tunisian Islamist Sites", URL:
http://www.google.com/hostednews/afp/article/ALeqM5hwgwVBJdpJa0Jm1LsoCpW0HnophQ?docId=CNG.a7aa7a5a7dd45ef26c7fad8a1c0a0dfe.401

[56] Al Jazeera, 19 May 2011, "Anonymous and The Arab Uprisings", URL:
http://www.aljazeera.com/news/middleeast/2011/05/201151917634659824.html

[57] Naked Security, 26 Jan. 2011, "Egypt versus the Internet – Anonymous Hackers Launch DDoS-Attack", URL: http://nakedsecurity.sophos.com/2011/01/26/egypt-versus-the-internet-anonymous-hackers-launch-ddos-attack/

[58] Naked Security, 26 Jan. 2011, "Egypt versus the Internet – Anonymous Hackers Launch DDoS-Attack", URL: http://nakedsecurity.sophos.com/2011/01/26/egypt-versus-the-internet-anonymous-hackers-launch-ddos-attack/

[59] Al Jazeera, 19 May 2011, "Anonymous and The Arab Uprisings", URL:
http://www.aljazeera.com/news/middleeast/2011/05/201151917634659824.html

[60] Huffington Post, 8 Aug. 2011, "Syrian Ministry of Defense Website Hacked by Anonymous", URL:
http://www.huffingtonpost.com/2011/08/08/syria-ministry-of-defense-hacked-anonymous_n_920733.html

[61] Gawker, 2 Feb. 2011, "Anonymous Hackerse Attack Yemeni Government", URL:
http://gawker.com/5750513/anonymous-hackers-already-taking-down-yemeni-websites

[62] Softpedia, 17 July 2012, "Anonymous Hackers Publish Details of Yemen's Internet Filtering Systems", URL: http://news.softpedia.com/news/Anonymous-Hackers-Publish-Details-of-Yemen-s-Internet-Filtering-Systems-281745.shtml

[63] The Hacker News, 11 Feb. 2011, "Operation Algeria, Part 2 by Anonymous Hackers Released", URL: http://thehackernews.com/2011/02/operation-algeria-part-2-by-anonymous.html

[64] Naked Security, 31 Dec. 2010, "Pro-WikiLeaks Hackers Attack Zimbabwe Government Websites", URL: http://nakedsecurity.sophos.com/2010/12/31/pro-wikileaks-hackers-attack-zimbabwe-government-websites/

[65] The Hacker News, 10 Feb. 2011, "Operation Italy Press Release: Anonymous Hackers Will Soon Strike Again", URL: http://thehackernews.com/2011/02/operation-italy-press-release-anonymous.html

[66] Daily Caller, 26 Feb. 2011, "Who is Anonymous? A Look At the Hacktivists Aiding Revolution in the Middle East", URL: http://dailycaller.com/2011/02/26/who-is-anonymous-a-look-at-the-hacktivists-aiding-revolution-in-the-middle-east/2/

[67] Al Jazeera, 19 May 2011, "Anonymous and The Arab Uprisings", URL: http://www.aljazeera.com/news/middleeast/2011/05/201151917634659824.html

[68] Ars Technica, 15 Feb 2011, "Anonymous Speaks: The Inside Story of the HBGary Hack", URL: http://arstechnica.com/tech-policy/2011/02/anonymous-speaks-the-inside-story-of-the-hbgary-hack/

[69] Olson, Parmy, 2012, "We Are Anonymous – Inside the Hacker World of LulzSec, Anonymous, And the Global Cyber Insurgency", Little Brown, pp.17

[70] eSecurity, 8 Feb. 2011, "Anonymous Hackers Target HB Gary", URL: http://www.esecurityplanet.com/headlines/article.php/3923906/Anonymous-Hackers-Target-HBGary.htm

[71] Ars Technica, 15 Feb 2011, "Anonymous Speaks: The Inside Story of the HBGary Hack", URL: http://arstechnica.com/tech-policy/2011/02/anonymous-speaks-the-inside-story-of-the-hbgary-hack/

[72] Ars Technica, 15 Feb 2011, "Anonymous Speaks: The Inside Story of the HBGary Hack", URL: http://arstechnica.com/tech-policy/2011/02/anonymous-speaks-the-inside-story-of-the-hbgary-hack/

[73] Olson, Parmy, 2012, "We Are Anonymous – Inside the Hacker World of LulzSec, Anonymous, And the Global Cyber Insurgency", Little Brown, p. 20

[74] Ars Technica, 15 Feb 2011, "Anonymous Speaks: The Inside Story of the HBGary Hack", URL: http://arstechnica.com/tech-policy/2011/02/anonymous-speaks-the-inside-story-of-the-hbgary-hack/

[75] Ibid.

Excerpt from *Introduction to Cyber-Warfare: A Multidisciplinary Approach*
*(Shakarian, Shakarian, and Ruef)*

[76] Ars Technica, 15 Feb 2011, "Anonymous Speaks: The Inside Story of the HBGary Hack", URL:
http://arstechnica.com/tech-policy/2011/02/anonymous-speaks-the-inside-story-of-the-hbgary-hack/

[77] eSecurity, 8 Feb. 2011, "Anonymous Hackers Target HB Gary", URL:
http://www.esecurityplanet.com/headlines/article.php/3923906/Anonymous-Hackers-Target-HBGary.htm

[78] Ars Technica, 4 Apr. 2011, "'Anonymous' Attacks Sony to Protest PS3 Hacker Law Suit", URL:
http://arstechnica.com/tech-policy/2011/04/anonymous-attacks-sony-to-protest-ps3-hacker-lawsuit/

[79] New Yorker, 7 May 2012, "Machine Politics – Annals of Technology", URL:
http://www.newyorker.com/reporting/2012/05/07/120507fa_fact_kushner

[80] Ps3News, undated, "PS3 Is Hacked By George Hotz – Hello Hypervisor, I'm GeoHot!", URL:
http://www.ps3news.com/PS3-Hacks/ps3-is-hacked-by-george-hotz-hello-hypervisor-im-geohot/

[81] Technologizer, 11 Apr. 2011, "Sony And George Hotz Settle PS3 Hacking Lawsuit", URL:
http://technologizer.com/2011/04/11/sony-george-hotz-settle-ps3-hacking-lawsuit/

[82] BusinessInsider, 24 Jan. 2012, "Famous iPhone Hacker George Hotz Has Left Facebook", URL:
http://articles.businessinsider.com/2012-01-24/tech/30658124_1_george-hotz-location-data-iphone

[83] Olson, Parmy, 2012, "We Are Anonymous – Inside the Hacker World of LulzSec, Anonymous, And the Global Cyber Insurgency", Little Brown, pp. 423f

[84] Know Your Meme, 1 March 2012, "Operation Antisec", URL:
http://knowyourmeme.com/memes/events/operation-antisec

[85] PCMagazine, 20 June 2011, "LulzSec, Anonymous Team Up for 'Operation Anti-Security", URL:
http://www.pcmag.com/article2/0,2817,2387264,00.asp

[86] PCMagazine, 20 June 2011, "LulzSec, Anonymous Team Up for 'Operation Anti-Security", URL:
http://www.pcmag.com/article2/0,2817,2387264,00.asp

[87] Know Your Meme, 1 March 2012, "Operation Antisec", URL:
http://knowyourmeme.com/memes/events/operation-antisec

[88] ITProPortal, 27 June 2011, "Anonymous Puts US Counter Terrorist Program Online", URL:
http://www.itproportal.com/2011/06/27/anonymous-puts-us-counter-terrorist-program-online/

[89] Know Your Meme, 1 March 2012, "Operation Antisec", URL: http://knowyourmeme.com/memes/events/operation-antisec

[90] The Telegraph, 7 March 2012, "FBI Charges Alleged Anonymous Hackers After Supergrass Claims", URL: http://www.telegraph.co.uk/technology/news/9127004/FBI-charges-alleged-Anonymous-hackers-after-supergrass-claims.html

[91] Peoples Liberation Front, Public Campaigns, undated, "Operation Vendetta", URL: http://www.peoplesliberationfront.net/campaignsArchive1.html, last accessed 08/12/2012

[92] TPM, 8 March 2012, "Occupy Lawyer Stuck With $35K Bill After 'Homeless Hacker' Jumps Bail", URL: http://idealab.talkingpointsmemo.com/2012/03/occupy-lawyer-stuck-with-35k-bill-after-homeless-hacker-jumps-bail.php

[93] TPM, 27 Oct. 2011, "'Homeless Hacker' Christopher Doyon, AKA 'Commander X', Joins Up With Occupy Movement", URL: http://idealab.talkingpointsmemo.com/2011/10/homeless-hacker-christopher-doyon-aka-commander-x-joins-up-with-occupy-movement.php

[94] CBS News, 23 Sep. 2011, "Feds: Homeless Hacker 'Commander X' Arrested", URL: http://www.cbsnews.com/8301-31727_162-20110912-10391695.html

[95] The Montreal Gazette, 12 May 2012, "Christopher Doyon, One of the Brains Behind Anonymous", URL: http://www.montrealgazette.com/technology/Christopher+Doyon+brains+behind+Anonymous/6612381/story.html

[96] Wired, 13 July 2012, "Par:AnoIA: Anonymous Launches WikiLeaks-esque Site for Data Dumps", URL: http://www.wired.com/threatlevel/2012/07/paranoia-anonymous/all/

[97] Wired, 19 Apr 2012, "Anontune: The New Social Music Platform From Anonymous", URL: http://www.wired.com/underwire/2012/04/anontune-anonymous/

[98] RT.com, 23 Apr 2012, "Anonymous Unveils Data-Sharing Websites Amidst Privacy Concerns", URL: http://rt.com/usa/news/anonymous-pastebin-new-internet-605/

[99] RT.com, 23 Apr 2012, "Anonymous Unveils Data-Sharing Websites Amidst Privacy Concerns", URL: http://rt.com/usa/news/anonymous-pastebin-new-internet-605/

[100] RT.com, 23 Apr 2012, "Anonymous Unveils Data-Sharing Websites Amidst Privacy Concerns", URL: http://rt.com/usa/news/anonymous-pastebin-new-internet-605/

[101] Information Week, 19 April 2012, "Anonymous Builds New Haven for Stolen Data", URL: http://www.informationweek.com/news/security/vulnerabilities/232900590

[102] PC Magazine, 19 Apr. 2012, "With PasteBin 'Compromised', Anonymous Launches AnonPaste", URL: http://www.pcmag.com/article2/0,2817,2403283,00.asp

[103] Wired, 13 July 2012, "Par:AnoIA: Anonymous Launches WikiLeaks-esque Site for Data Dumps", URL: http://www.wired.com/threatlevel/2012/07/paranoia-anonymous/all/

[104] Wired, 13 July 2012, "Par:AnoIA: Anonymous Launches WikiLeaks-esque Site for Data Dumps", URL: http://www.wired.com/threatlevel/2012/07/paranoia-anonymous/all/

[105] CNet, 15 March 2012, "Anonymous OS: Worth The Risk?", URL: http://news.cnet.com/8301-13506_3-57397895-17/anonymous-os-worth-the-risk/

[106] Mobiledia, 16 March 2012, "Hackers' OS? Joke's On You", URL: http://www.mobiledia.com/news/133380.html

[107] The Next Web (TNW), undated, "Anonymous Has Just Released Its Own Operating System: Anonymous OS", URL: http://thenextweb.com/insider/2012/03/14/anonymous-has-just-released-its-own-operating-system-anonymous-os/

[108] The Hacker News, 15 March 2012, "Anonymous-OS 0.1: Anonymous Hackers Released Their Own Operating System", URL: http://thehackernews.com/2012/03/anonymous-os-01-anonymous-hackers.html

[109] CNet, 15 March 2012, "Anonymous OS: Worth The Risk?", URL: http://news.cnet.com/8301-13506_3-57397895-17/anonymous-os-worth-the-risk/

[110] The Next Web (TNW), undated, "Anonymous Has Just Released Its Own Operating System: Anonymous OS", URL: http://thenextweb.com/insider/2012/03/14/anonymous-has-just-released-its-own-operating-system-anonymous-os/

[111] Softpedia, 16 March 2012, "SourceForge Closes Anonymous-OS Live CD Project", URL: http://news.softpedia.com/newsImage/SourceForge-Closes-Anonymous-OS-Live-CD-Project-2.jpg/

[112] The Next Web, undated, "Anonymous Claims That The Operating System, 'Anonymous-OS' Is Fake", URL: http://thenextweb.com/insider/2012/03/15/anonymous-claims-that-the-operating-system-anonymous-os-is-fake/

Excerpt from *Introduction to Cyber-Warfare: A Multidisciplinary Approach*
*(Shakarian, Shakarian, and Ruef)*

[113] Softpedia, 16 March 2012, "SourceForge Closes Anonymous-OS Live CD Project", URL: http://news.softpedia.com/news/SourceForge-Closes-Anonymous-OS-Live-CD-Project-258944.shtml

[114] The Next Web, undated, "Anonymous Claims That The Operating System, 'Anonymous-OS' Is Fake", URL: http://thenextweb.com/insider/2012/03/15/anonymous-claims-that-the-operating-system-anonymous-os-is-fake/

[115] Mobiledia, 16 March 2012, "Hackers' OS? Joke's On You", URL: http://www.mobiledia.com/news/133380.html

[116] The Next Web, undated, "Anonymous Claims That The Operating System, 'Anonymous-OS' Is Fake", URL: http://thenextweb.com/insider/2012/03/15/anonymous-claims-that-the-operating-system-anonymous-os-is-fake/

[117] ZDNet, 16 July 2012, "Fluid Structure, Leadership Keep Anonymous' Threat Alive", URL: http://www.zdnet.com/fluid-structure-leadership-keep-anonymous-threat-alive-7000000957/

[118] Radio Free Europe, 3 Mar 2012, "What Is Anonymous And How Does It Operate?", URL: http://www.rferl.org/content/explainer_what_is_anonymous_and_how_does_it_operate/24500381.html

[119] Mobiledia, 24 Apr. 2012, "A Rising War Between Hackers", URL: http://www.mobiledia.com/news/139185.html